\documentclass[graybox]{svmult}

\usepackage{mathptmx}       
\usepackage{helvet}         
\usepackage{courier}        
\usepackage{type1cm}        
\usepackage{makeidx}         
\usepackage{graphicx}        
\usepackage{multicol}        
\usepackage[bottom]{footmisc}

\makeindex             

\begin{document}

\title*{Mean-field transport of a Bose-Einstein condensate}
\author{Samy Mailoud Sekkouri and Sandro Wimberger}
\institute{Samy Mailoud Sekkouri \at DiFeST, Universit\`a degli Studi di Parma, Via G. P. Usberti 7/a, 43124 Parma, Italy
\and  Sandro Wimberger \at DiFeST, Universit\`a degli Studi di Parma, Via G. P. Usberti 7/a, 43124 Parma, Italy and INFN, Sezione di Milano Bicocca, Gruppo Collegato di Parma, \email{sandro.wimberger@fis.unipr.it}}
%
%
\maketitle


\abstract{
The expansion of an initially confined Bose-Einstein condensate into either free space or a tilted optical lattice is investigated in a mean-field approach. The effect of the interactions is to enhance or suppress the transport depending on the sign and strength of the interactions. These effects are discusses in detail in view of recent experiments probing non-equilibrium transport of ultracold quantum gases.
}

\section{Introduction}
\label{sec:1}

Since the first realization of Bose-Einstein condensates in 1995 with ultracold alkali atoms \cite{BEC}, experiments with ultracold quantum gases have launched a vast research field for investigating the quantum nature of matter with an unprecedented experimental precision \cite{BDZ,Weidemueller}. One of the directions investigated today is concerned with the quantum transport of ultracold matter. Pioneering here are the recent experimental results by the two groups at ETH \cite{ETH} and at NIST \cite{NIST}. Many transport scenarios of ultracold bosons and fermions were studied starting from a microscopic (many-body) description \cite{ManyB,Ivan13}. In a more general setting, mean-field quantum transport of a Bose-Einstein condensate was investigated in the context of Bloch oscillations and tunneling in Wannier-Stark systems \cite{Pisa,WSTheo}, of barrier tunneling \cite{Regensburg}, of disorder, \cite{Florence}, or of time-dependent potentials \cite{Floquet}. In almost all of the experimental realizations, so far, what has been studied was essentially the expansion of a cloud of cold atoms which is controlled by external fields and interactions. Along the same lines, we propose in this contribution a relatively simple method to prepare the initial state, namely within an steep harmonic trap. Transport occurs when the trap is opened in one direction. We investigate in detail how the particle current in such a setup depends on the interactions, which we treat in mean-field approximation following the celebrated Gross-Pitaevskii equation \cite{Leggett,Books}.

\section{Out transport setup}
\label{sec:2}

The dynamics of a Bose-Einstein condensate in mean-field approximation is described by the Hamiltonian
\begin{equation}
\label{eq:1}
	H= \frac{p^2}{2m} + V_{\rm int} (r,t) + V_{\rm ext} (r,t) \,.
\end{equation}
The interatomic potential of a cold dilute gas of bosons is replaced by the effective mean-field interaction
\begin{equation}
	V_{\rm int} (r,t)= g_{3D} |\psi (r,t)|^2 \,,
\end{equation}
where the coupling constant $g_{3D}= \frac{4 \pi \hbar a_s}{M} N$ is determined by the number of atoms $N$, their mass $M$, and the two-body s-wave scattering length $a_s$. The wave function is then normalized to unity. Please note that the strength and the sign of $a_s$ can be controlled quite well in the experiment \cite{BDZ,Books}. We restrict here to a quasi one-dimensional situation, in which the condensate is well confined in the two transverse directions. Such a reduction essentially leads to a rescaling of the coupling constant. This rescaling depends on the precise geometry of the trapping potentials. A standard argument \cite{BDZ} reduces $g$ to its one-dimensional version $g_{1D}= 2 \hbar \omega_\perp a_s$, where the transverse confinement frequency $\omega_\perp$ is assumed to be large compared to the longitudinal one.
\begin{figure}[b]
\begin{center}(a)
\includegraphics[width=0.75\textwidth]{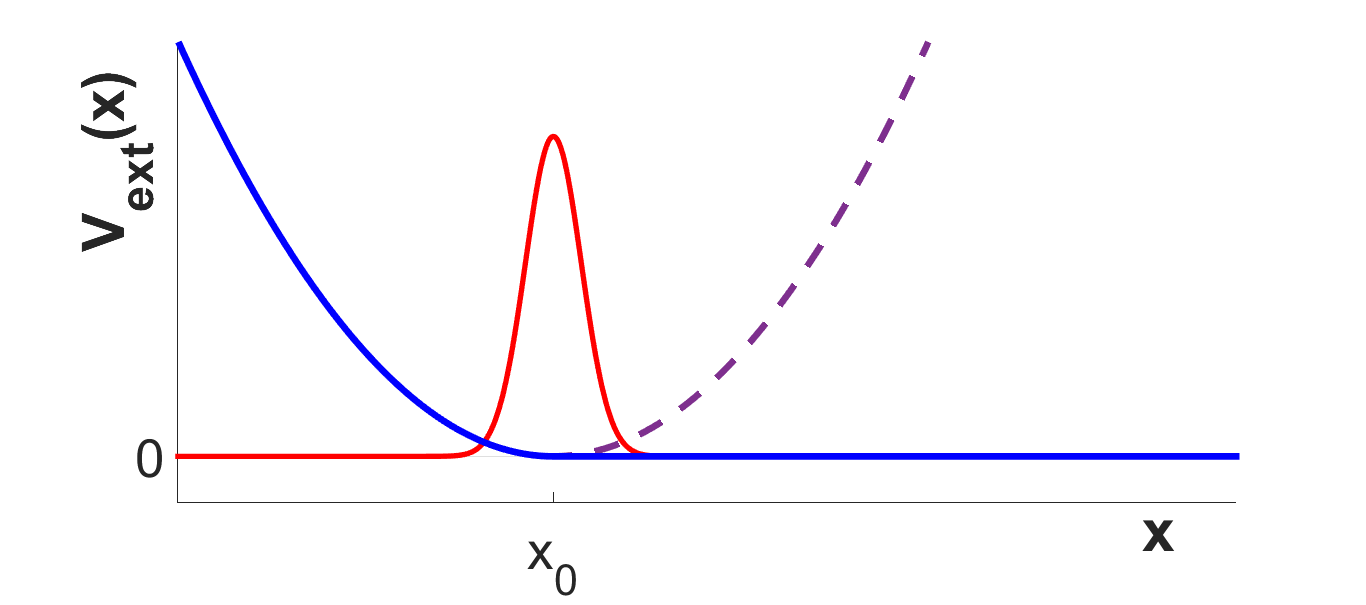} \\
(b)
\includegraphics[width=0.75\textwidth]{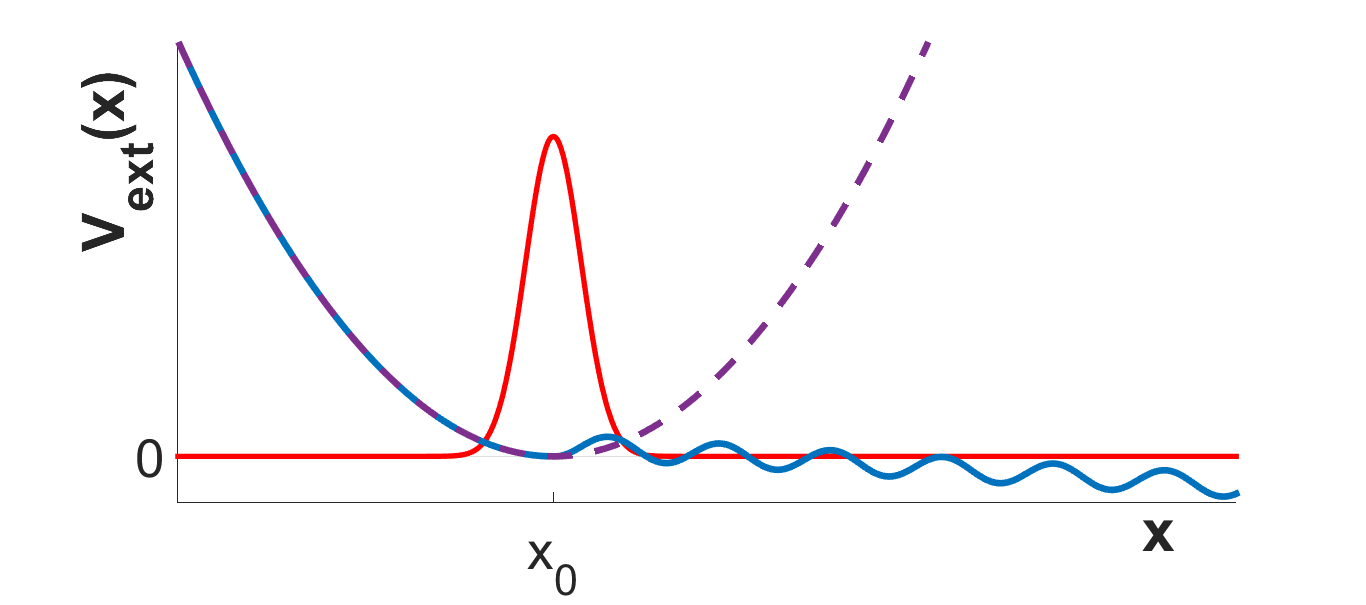}
\end{center}
\caption{Sketch of the experiments we are proposing. The initial state (red solid lines) is prepared within an harmonic trap (blue lines for $x<x_0$ and viola dashed lines for $x>x_0$). The trap is released on the right part of $x_0$ to zero, which makes the initial wavepacket move towards the right. We investigate two exemplary cases: without any external potential seen in (a) and with a tilted periodic lattice seen in (b). The total external potential at $t>0$ is plotted by the overall blue lines in both cases.}
\label{fig:1}       
\end{figure}

To simplify the problem, we express all quantities in the units of the longitudinal harmonic oscillator confinement with frequency $\omega$ at $t=0$. This means that we express $p\to\tilde{p}\equiv p (\hbar \omega m)^{-\frac{1}{2}}$ and $x\to\tilde{x}\equiv x \left(\frac{\omega m}{\hbar}\right)^{\frac{1}{2}}$. In this units, the Hamiltonian for $t=0$ now reads
\begin{equation}
\label{eq:ham1}
\tilde{H}(\tilde{x},\tilde{p};t=0) \ = \ \frac{1}{2} \tilde{p}^2 \ + \ \tilde{g}_{1D} |\tilde{\psi} (\tilde{x},\tilde{t})|^2   
 \ + \ \frac{1}{2} (\tilde{x}-\tilde{x}_0 )^2  \,,
\end{equation}
and for $t>0$ correspondingly
\begin{eqnarray}
\label{eq:ham2}
\tilde{H}(\tilde{x},\tilde{p};t>0) & \ = \ & \frac{1}{2} \tilde{p}^2 \ + \  \tilde{g}_{1D} |\tilde{\psi} (\tilde{x},\tilde{t})|^2   
 + \ \frac{1}{2} \Theta(\tilde{x}_0-\tilde{x})(\tilde{x}-\tilde{x}_0 )^2  \\
& - &\ \Theta (\tilde{x}-\tilde{x}_0)\left[\tilde{F} \tilde{x} + \tilde{A} \sin^2(\tilde{K}(\tilde{x}-\tilde{x_0}))\right] \,.
\end{eqnarray}
The initially prepared state and the potentials are sketched in Fig. \ref{fig:1}. The sinusoidal term in Eq. (\ref{eq:ham2}) describes an optical lattice into which the condensate can expand. $\tilde{A}$ is the amplitude of the lattice and $\tilde{K}=\pi/\tilde d_{\rm L}$ determines its spatial period $\tilde d_{\rm L}$. The linear potential with force $\tilde{F}$  controls the tilt of the lattice. In the next section we study the temporal evolution in the sketched setups, in particular the dependence of the atomic currents (towards the right) on the interaction strength $\tilde{g}_{1D}$. In the following we drop the tildes for simplicity, with the additional convention $g \equiv \tilde{g}_{1D}$.  

\section{Numerical results}
\label{sec:3}

\begin{figure}[t]
\begin{center}
\includegraphics[width=\textwidth]{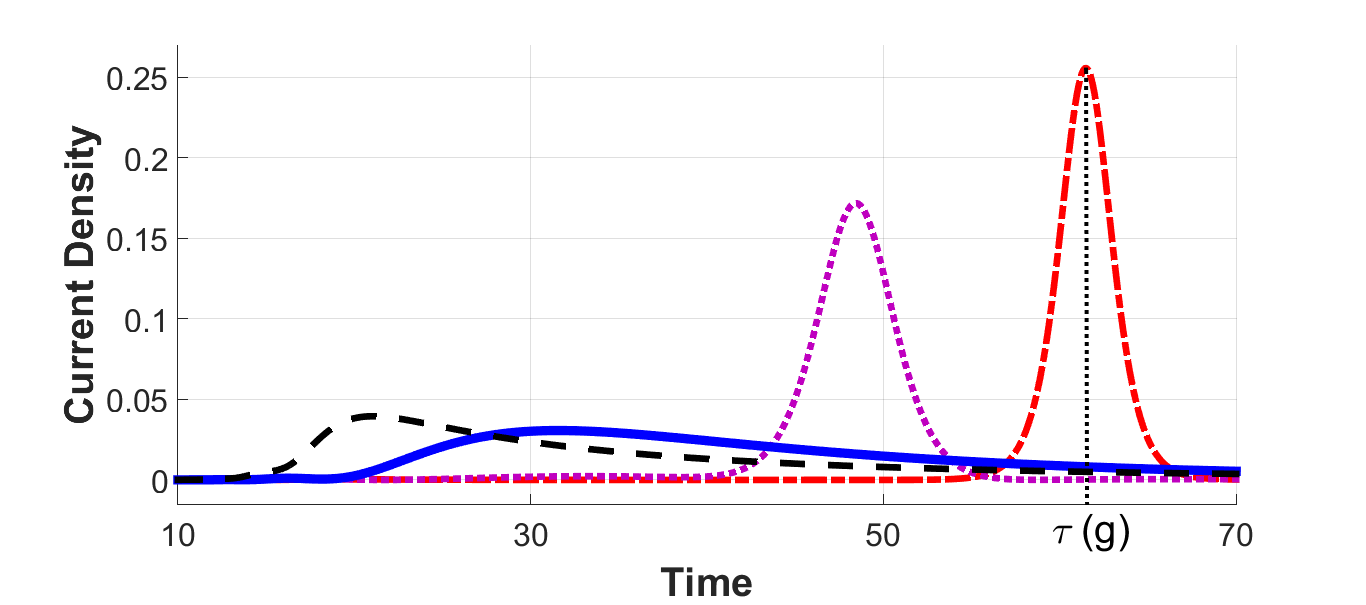} \\
\end{center}
\caption{The particle current as a function of time at $x=2x_0$ for the following values of interaction strength from left to right: $g=2$ (black dashed line), $g=0$ (blue solid line), $g=-1$ (viola dotted line), and $g=-2$ (red dot-dashed curve). We observe clear maxima of the currents, whose position on the time axis (denoted by $\tau$) is determined by the sign and the strength of the nonlinearity. The lattice parameters are $A=1, d_{\rm L}=4$ and $F=0.043$.
}
\label{fig:2}       
\end{figure}
Our main observable for the study of the mean-field transport of the condensate is the following probability current density 
\begin{equation}
\label{eq:cur1}
	j(x,t) = \frac{1}{2i} [\psi ^* (x,t) \frac{\partial \psi(x,t)}{\partial x}   - \psi(x,t) \frac{\partial \psi ^*(x,t)}{\partial x}  ] \,.
\end{equation}
The current will obviously depend on the precise nature of the interaction (attractive or repulsive) and its strength. We integrate the
nonlinear Schr\"odinger equation determined by Eq. (\ref{eq:ham2}), using a finite difference spatial representation of the wave function and a norm-preserving Crank-Nicholson integrator, see e.g. \cite{CN} for details on the integration scheme. For a grid-step size of $\Delta x$, the time-dependent current at the grid point $x_n$ is given by
\begin{equation}
\label{eq:cur2}
	j(x_n,t) = \frac{i}{2\Delta x} [\psi ^* (x_n+\Delta x,t) \psi(x_n,t) - \psi^*(x_n,t) \psi (x_n+\Delta x,t) ]\,.
\end{equation}

\begin{figure}[t]
\begin{center}
\includegraphics[width=\textwidth]{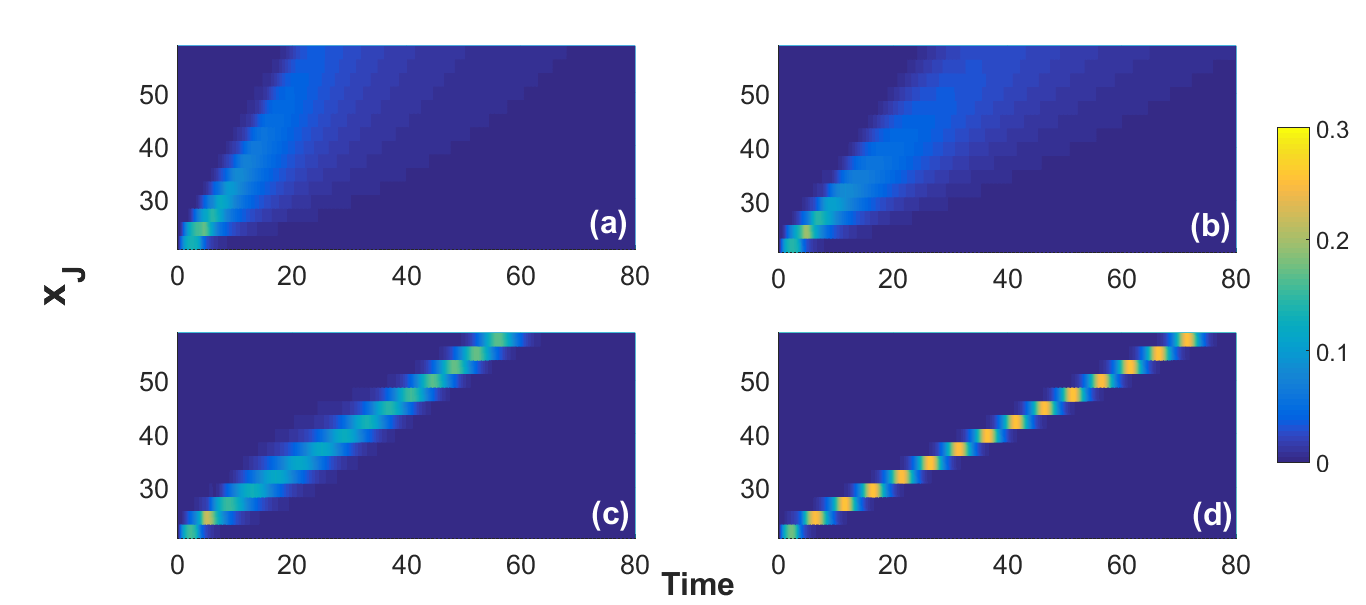} \\
\end{center}
\caption{Heat map of the current density $j(x,t)$ as in Fig. \ref{fig:1} but for a window of positions $x=x_J$. The two-dimensional plots show that repulsive interactions enhance the transport, see panel (a) for $g=2$, while attractive interactions slow it down, see panel (c) for $g=-1$ and (d) for $g=-2$. (b) is the reference case without interactions. As an interesting side effect, the dispersion in the spatial-temporal plane $(x,t)$ is minimized by strong attractive interactions, see panel (d), corresponding to the dot-dashed line in the previous figure.
}
\label{fig:2-2}       
\end{figure}

\subsection{Case (a): Directed free expansion}
\label{subsec:1}

\begin{figure}[t]
\begin{center}
\includegraphics[width=0.75\textwidth]{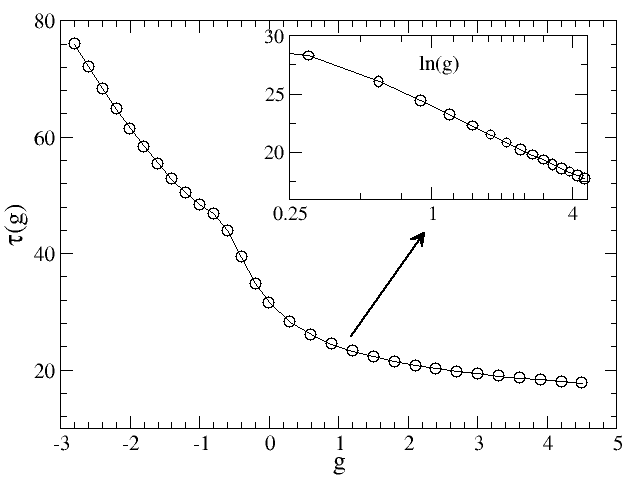} \\
\end{center}
\caption{The times $\tau (g)$ of maximal current at position $x=2x_0$ extracted from data sets such as shown in Fig. \ref{fig:2} and for the same lattice parameters as there. For positive nonlinearities $g$, the scaling of the enhancement of the expansion seems logarithmic (see inset). For negative $g$, the expansion is slowed down a lot, which can be seen by the steep increase of the curve for decreasing $g<0$. 
}
\label{fig:3}       
\end{figure}

For the case of the free expansion towards the right (case (a) in Fig. \ref{fig:1}), we first plot the current density as a function of the interaction strength $g$ at the point $x=2x_0$, with $x_0=20.5$, please see Fig. \ref{fig:2}. At this fixed position, the probability current as a function of time shows a characteristic maximum, whose position on the time axis is determined by $g$. 

For repulsive interactions ($g>0$), the wave packet tends to expand faster due to the additional repulsive potential term in Eq. (\ref{eq:ham1}). For the attractive case ($g<0$), the opposite happens and the wave packet tends to stabilize and the expansion is slowed down. Figure \ref{fig:2-2} shows the same results for a window of positions from $x=x_0$ to $x \approx 60$ (above which the wave function is absorbed in order to avoid artificial back reflections). Interestingly, but not too surprisingly, the dispersion in the spatial-temporal plane $(x,t)$ is minimized by strong attractive interactions. Here the current maximum is very stable and the dynamics of the condensate is almost free of dispersion similar to a solitonic motion, see Fig. \ref{fig:2-2}(d).

In order to quantify the effect of the nonlinearity $g$, we plot the dependence of the times $\tau$ when the maximum density is reached at $x=2x_0$ in Fig. \ref{fig:3}. While the qualitative behavior of the enhanced expansion and the slowdown for positive and negative $g$, respectively, is clear (see also \cite{Holl1996}), we have no analytic explanation so far for the precise form of the observed scaling of $\tau(g)$ seen in Fig. \ref{fig:3}.

\subsection{Case (b): Expansion into a Wannier-Stark lattice}
\label{subsec:2}

\begin{figure}[t]
\begin{center}
\includegraphics[width=0.9\textwidth]{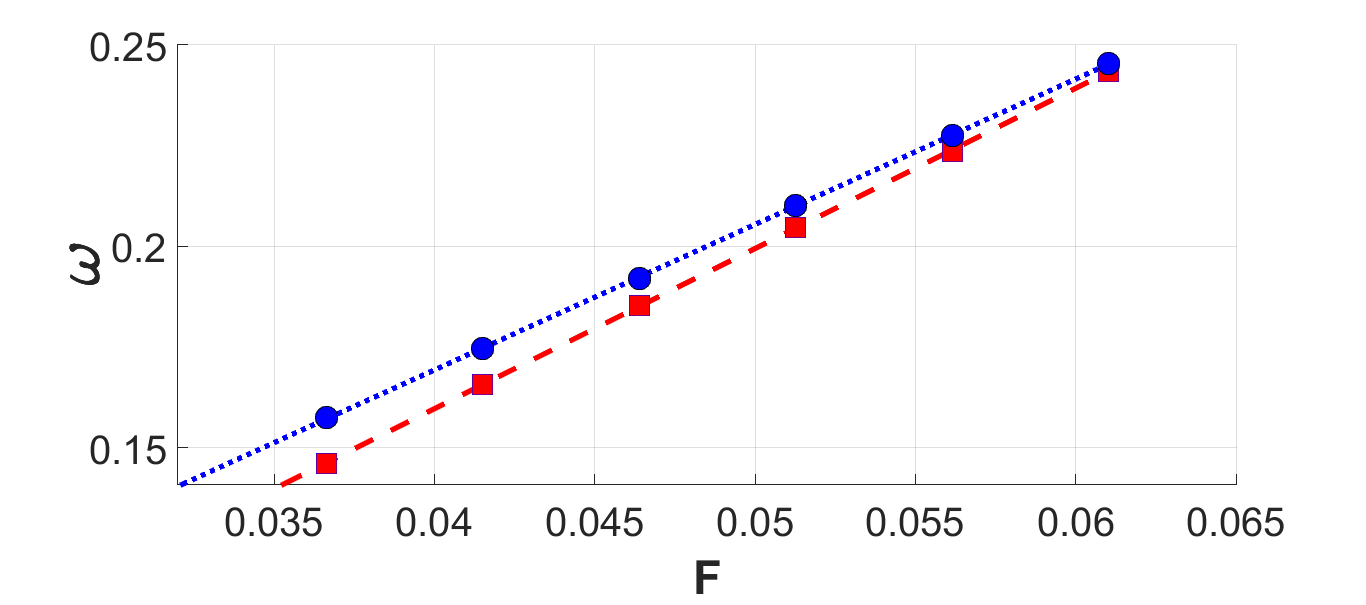} \\
\end{center}
\caption{Oscillation frequency $\omega$ at $g=0$ vs. the Stark force $F$ for the case with left confinement (blue symbols connected by dotted line) and without it (red symbols connected by dashed line). In both cases, the scaling is linear as expected. The presence of the left part of the harmonic trap affects only the slope. The lattice parameters are $A=1$ and $d_{\rm L}=4$.
}
\label{fig:4}       
\end{figure}

Optical lattices are by now a standard tool for the control of the motion of Bose-Einstein condensates \cite{Weidemueller}. The presence of an optical lattice slows down the expansion into it, while a constant negative tilt accelerates an initially localized wave packet towards the right. However, when both potential are present simultaneously, c.f. our setup shown in Fig. \ref{fig:1}(b), the situation is less clear. A tilted lattice problem defines the Wannier-Stark system, which was investigated with Bose condensates in great detail before, see e.g. \cite{Pisa,WSTheo}. In this system, again an initially localized wave packet remains localized but it oscillates with a characteristic Bloch frequency $\omega_{\rm B}$ given by the constant level distance of the energy spectrum (arising from the constant spatial tilt).  In our units, $\omega_{\rm B}=Fd_{\rm L}$, where $d_{\rm L}$ is the lattice spacing. This linear scaling of the oscillation frequency with the Stark force $F$ is seen also in our expansion problem in the absence of interactions ($g=0$). Because of the presence of the harmonic confinement on the left, the proportionality factor is slightly lower than one, as seen in Fig. \ref{fig:4} (blue symbols connected by the dotted line). Releasing also the left part of the trap, we instead observe the correct pre factor one, please see the red symbols in Fig. \ref{fig:4}. The frequencies are extracted from the current oscillations to the right of (but close to) $x_0$ after a short initial transient, necessary for the wave packet to adapt to the presence of the tilted lattice.

\begin{figure}[t]
\begin{center}(a)
\includegraphics[width=0.75\textwidth]{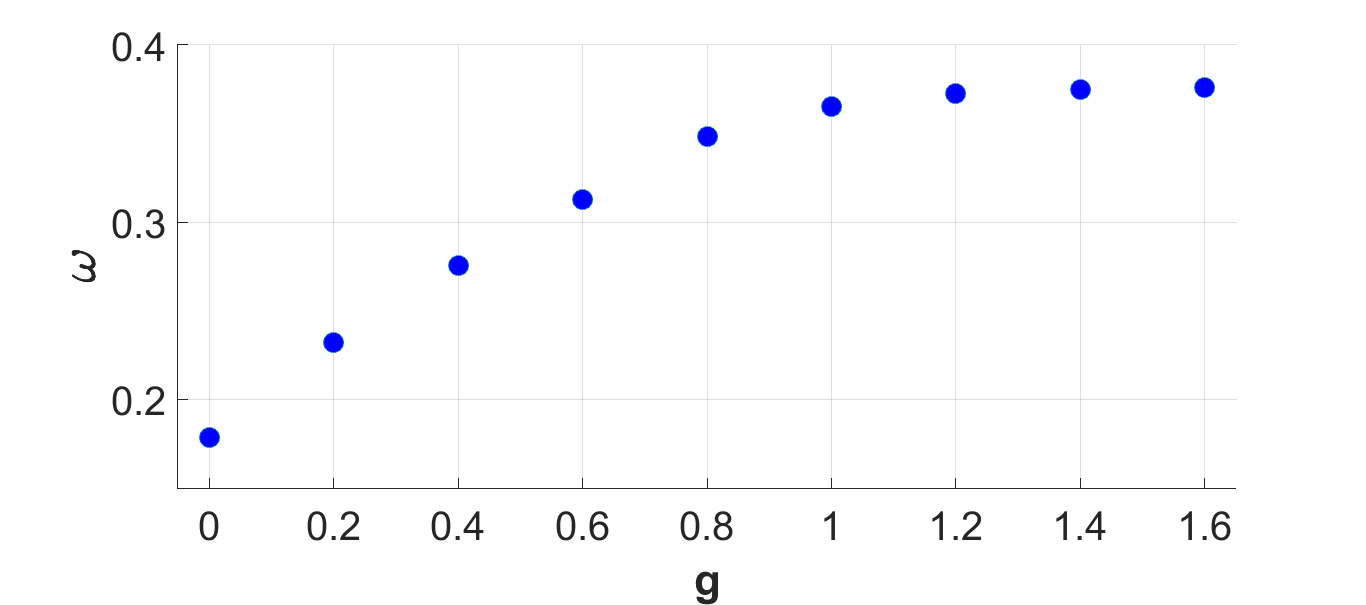} \\
(b)
\includegraphics[width=0.75\textwidth]{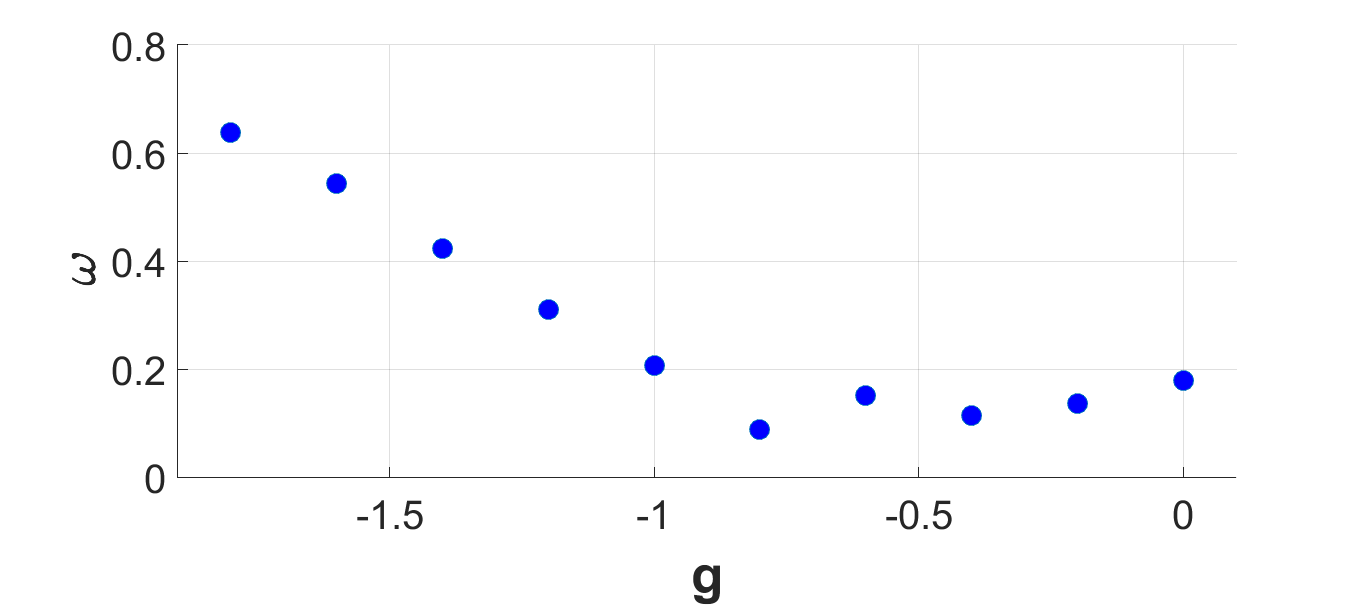}
\end{center}
\caption{Bloch-like oscillation frequency $\omega$ as a function of the nonlinear coupling constant $g$. While an attractive interaction ($g<0$, see (a)) slows down the oscillations in the region $-0.8 < g < 0$, a repulsive one, see (b), increases the frequency. For large negative $g<-1$, the nonlinear potential dominates the dynamics and interaction-induced oscillations with a frequency $\omega \propto |g|$ occur. The lattice parameters are the same as in the previous figure.}
\label{fig:5}       
\end{figure}

More interesting is the oscillatory behavior in the presence of interactions. We investigate again both cases of repulsive and attractive nonlinearity. The frequencies are extracted as described above from the current oscillations. Our results are shown in Fig. \ref{fig:5}. A repulsive interaction with $g>0$ increases the oscillations frequency. For not too large positive $g$, this increase is linear, and we will come up with an intuitive explanation below. For large nonlinearities a saturation is observed, see $g>1$ in Fig. \ref{fig:5}. Here the repulsion leads to a fast expansion which in turn decreases the density again. More complex is the case of attractive interactions with $g<0$. For small $|g|<1$, the Bloch-like oscillations are rather stable. For large $|g|>1$, again the nonlinearity potential dominates the dynamics, in the sense that the nonlinear term is larger than the kinetic term in Eq. (\ref{eq:ham2}). Here interaction-induced oscillations with a frequency $\omega \propto |g|$ occur. In this latter case, the density remains large during the evolution because of the attractive forces, and the theory developed by Kolovsky in \cite{Kolo2003} applies. There our observed linear scaling of the oscillation frequency with the nonlinear coupling parameter is theoretically predicted.

\begin{figure}[t]
\begin{center}(a)
\includegraphics[width=0.85\textwidth]{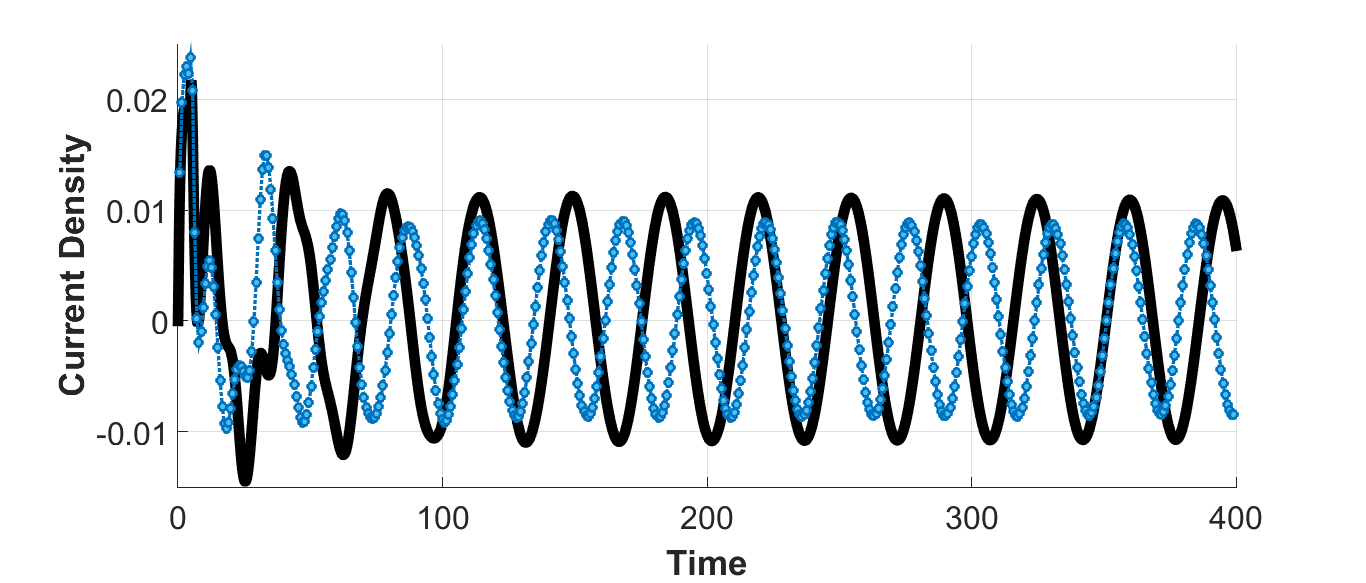} \\
(b)
\includegraphics[width=0.85\textwidth]{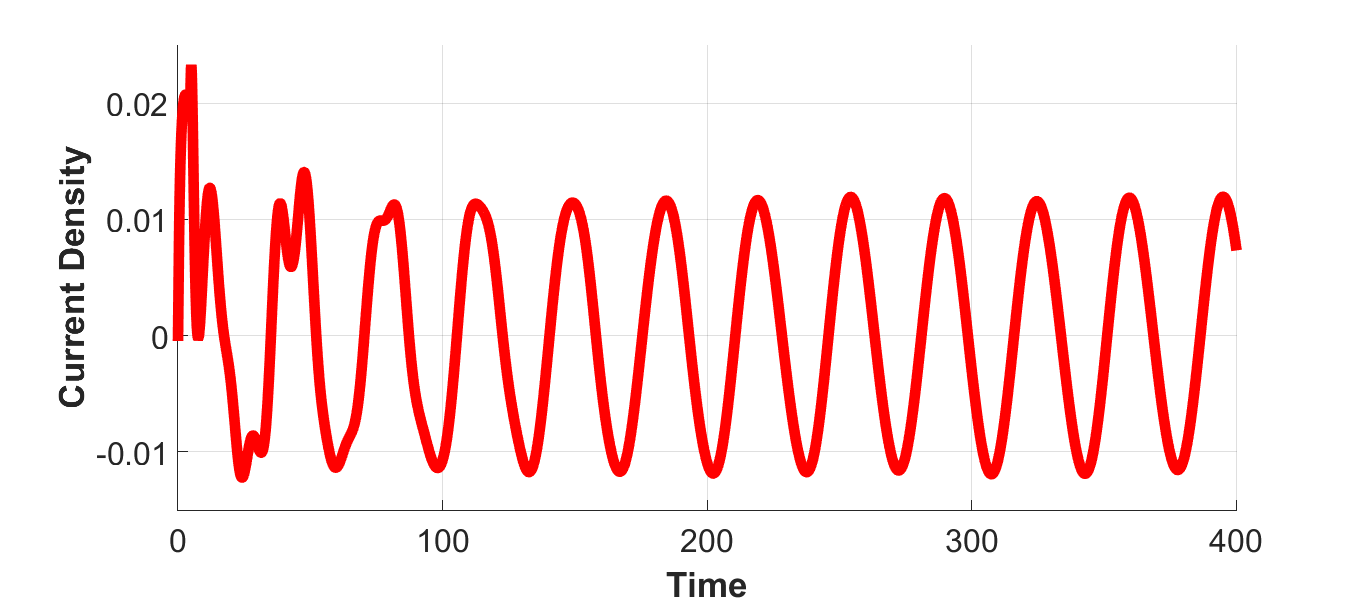}
\end{center}
\caption{Temporal evolution of the current to the right but close to $x_0$ for the three cases: (a) $g=0, F=0.043$ (black solid line), $g=0.2, F=0.043$ (blue symbols), and (b) $g=0.2, F=0.043-F'=0.029$ (red solid line). In (b) the nonlinear shift of the local energy level (where the atomic density is large) is corrected by a reduction of the Stark force with $F'=0.029$. We observe good agreement between the oscillation frequencies of the black (a) and the red (b) curves. The lattice parameters are chosen as in the previous two figures.
}
\label{fig:6}       
\end{figure}

\begin{figure}[t]
\begin{center}
\includegraphics[width=\textwidth]{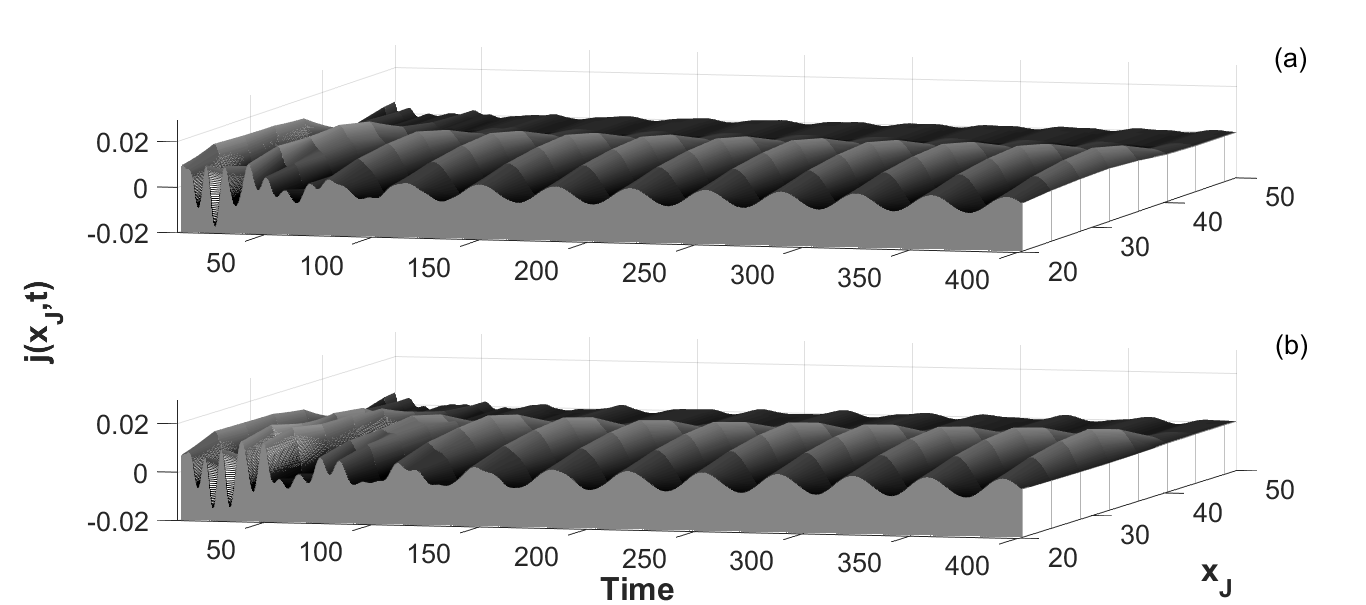}
\end{center}
\caption{Temporal evolution of the current as a function of position $x=x_J$ and time $t$. Shown are in panel (a) the data for the some parameters as in Fig. \ref{fig:6}(a) at $g=0$, and in panel (b) as in Fig. \ref{fig:6}(b). As noted previously the two cases are very similar due to the compensation of the effect of the nonlinearity in (b).
}
\label{fig:7}       
\end{figure}

In the following we concentrate on the case of repulsive interactions. Here we can explain the initial linear increase in the oscillation frequency seen Fig. \ref{fig:5}(b) by the local level shift induced by the nonlinear potential term in Eq. (\ref{eq:ham2}). This shift depends on the densities in the lattice sites which is largest in the ``central'' well centered at $x_0$ (at least at and close to $t=0$). This shift then leads to an effect increase of the difference $\Delta E$ of the two energy levels in the neighboring wells, and consequently to a larger oscillation frequency. We may estimate 
\begin{equation}
\label{eq:nonl}
	\Delta E \approx g \int_{d_{\rm L}} \ dx |\psi(x,t)|^2 \,.
\end{equation}
Because of the oscillations, we take the times $t$ of maximal density differences in the two wells for computing the above estimate. In principle, we can redo the effect of the nonlinear potential by rescaling the Stark force from $F$ to $F-F'$, where $F' \approx \Delta E / d_{\rm L}$. This reduces the problem to the noninteracting one with the same Block-like oscillation frequency determined just by $F$ alone. Corresponding numerical simulation for the current density are shown in Fig. \ref{fig:6} and Fig. \ref{fig:7}. The former plot nicely corroborates the effective compensation of the nonlinear potential in the temporal oscillations of the current. The latter figure highlights the good compensation comparing the currents globally in the spatial-temporal plane $(x>x_0,t)$.

Of course, our estimate given in Eq. (\ref{eq:nonl}) is a bit too rough in order to be perfect for all times (in particular because of the time-dependence of the process). Yet, this possibility of controlling the dynamics of a Bose-Einstein condensate is quite interesting. We refer to similar situations where the effect of the interaction was approximately cancelled by applying appropriate external potentials in theory \cite{Palermo} and an actual experiment at Innsbruck \cite{Innsbruck}. 

\section{Conclusions and perspectives}
\label{sec:4}

We propose a rather simple experiment to probe the effect of interparticle interactions in the non-equilibrium dynamics of a Bose-Einstein condensate. We have seen that the time-dependent atomic current towards the right can be well controlled in our setup. Interactions enhance or suppress the transport or the oscillations depending on their sign and their strength.

Preliminary computations on a full three dimensional evolution with strong confinement in the transverse dimensions seem to confirm our one-dimensional results (provided that the geometry of the confinement is matched such as to guarantee the same effective nonlinearity along the longitudinal direction). Interesting would be the case of an effective two-dimensional problem under so-called pancake confinement as recently studied in the context of mean-field transport in Kaiserslautern \cite{ott2015}. Here both directions are equally important and the expansion and transport of the condensate may be controlled even along both dimensions simultaneously.

\begin{acknowledgement}
We are very grateful for support by the FIL 2014 program of Parma University. Moreover, SW thanks the organizers of the NDES 2015 conference for their kind invitation and the wonderful meeting at Como.
\end{acknowledgement}

\end{document}